# Two-Dimensional Hyperferroelectric Metals: A Different Route to Ferromagnetic-Ferroelectric Multiferroics

Wei Luo[1,2], Ke Xu[1,2,3], and Hongjun Xiang[1,2]*

[1]*Key Laboratory of Computational Physical Sciences (Ministry of Education), State Key Laboratory of Surface Physics, and Department of Physics, Fudan University, Shanghai 200433, P. R. China*

[2]*Collaborative Innovation Center of Advanced Microstructures, Nanjing 210093, P. R. China*

[3]*Hubei Key Laboratory of Low Dimensional Optoelectronic Materials and Devices, Hubei University of Arts and Science, Xiangyang, 441053, P. R. China*

Email: hxiang@fudan.edu.cn

**Abstract:** Recently, two-dimensional (2D) multiferroics have attracted numerous attention due to their fascinating properties and promising applications. Although the ferroelectric (FE)-ferroelastic and ferromagnetic (FM)-ferroelastic multiferroics have been observed/predicted in 2D systems, 2D ferromagnetic-ferroelectric (FM-FE) multiferroics remain to be discovered since FM insulators are very rare. Here, we proposed for the first time the concept of 2D hyperferroelectric metals, with which the insulating prerequisite for the FM-FE multiferroic is no longer required in 2D systems. We validate the concept of 2D hyperferroelectric metals and 2D metallic FM-FE multiferroics by performing first-principle calculations on 2D CrN and $CrB_2$ systems. The 2D buckled monolayer CrN is found to be a hyperferroelectric metal with the FM ground state, i.e., a 2D FM-FE multiferroic. With the global optimization approach, we find the 2D $CrB_2$ system has an antiferromagnetic (AFM)/planar ground state and a FM/FE metastable state, suggesting that it can be used to realize electric field control of magnetism. Our analysis demonstrates that the spin-phonon coupling and metal-metal interaction are two new mechanisms for stabilizing the out-of-plane electric polarization in 2D systems. Our work not only extends the concept of FE to metallic systems, but also paves a new way to search the long-sought high temperature FM-FE multiferroics.

*Introduction*-Since graphene was experimentally discovered in 2004 [1], 2D materials (e.g., few-layer black phosphorus [2] and transition metal dichalcogenides [3,4]) have become an emerging topic area. These 2D materials are promising ingredients for developing high-speed-low power nanoelectronic devices. Very recently, 2D ferroelectricity [5-12], ferroelasticity [13,14] and multiferroicity [15-18] began to arouse interest due to the underlying new physics and promising applications as sensors, actuators, and memories [19-21]. For 2D in-plane ferroelectricity, experiment showed that SnTe thin-film (several unit-cells) is FE with the transition temperature higher than that of the bulk phase [22]. Other 2D SnTe analogues *MX* (*M*=Ge, Sn; *X*=S, Se) [6,15,18] were also predicted to exhibit in-plane ferroelectricity. Another calculation suggests that the covalently functionalized 2D materials with an in-plane ferroelectricity and a high-mobility can be used as FE field-effect transistors [23]. For 2D out-of-plane ferroelectricity, it is usually believed that it is unstable due to the presence of a depolarization field [24-26]. Nevertheless, some mechanisms were proposed to stabilize the 2D out-of-plane ferroelectricity. It was predicted that $MoS_2$ might be FE with an out-of-plane electric polarization due to the improper nature of the ferroelectricity [27]. The functionalized MXenes [28] also maintain out-of-plane polarization due to covalent interactions. In addition, the AB (e.g., GaAs) binary monolayers and 2D $P_2O_3$ maintain the proper out-of-plane electric polarization due to the presence of lone pair electrons [29,30], which can be seen as the 2D counterpart of insulating hyperferroelectric, i.e., a new class of proper ferroelectrics which polarize even when the depolarization field is unscreened [31,32].

For 2D multiferroics, it was proposed that ferroelectricity and ferroelasticity [15], or ferromagnetism and ferroelasticity [33] can coexist in 2D systems. For technological applications, it is highly desirable to combine FM and FE order within one material (i.e., FM-FE multiferroic) since FM-FE multiferroics [34-41] are ideal candidates to realize novel electric-write/magnetic-read memory devices. However, ferromagnetism is often incompatible with ferroelectricity since the ferromagnetism usually requires partially filled d-orbitals, while usual displacive ferroelectricity requires empty d-orbitals [42]. Furthermore, the usual FE is insulating, while the common strong ferromagnets are metallic. It remains a challenge to realize 2D FM-FE multiferroics.

In this article, we propose for the first time the concept of 2D hyperferroelectric metals and demonstrate that 2D hyperferroelectric metals can be adopted to design 2D FM-FE multiferroics. The metallicity of 2D hyperferroelectric metals is vital to host strong ferromagnetism. Meanwhile, the out-of-plane polarization of a 2D hyperferroelectric metal can be switched by a perpendicular electric field since the current is confined to flow within the 2D plane. With two new mechanisms (i.e., spin-phonon coupling and metal-metal bonding) for stabilizing the out-of-plane polarization in 2D hyperferroelectric metals, we predict that 2D CrN and $CrB_2$ are 2D magnetoelectric multiferroics.

*Computational details-*Our density functional theory (DFT) calculations are carried out using the Vienna ab initio simulation package (VASP) [43]. The projector augmented wave (PAW) method for treating ion-electron interactions [44,45] is used. The exchange-correlation potential is treated by the generalized gradient approximation (GGA) in the Perdew-Burke-Ernzerhof (PBE) form [46]. For the optimized structures, the atomic forces are less than 0.01 eV/Å. To avoid interactions between different layers, we set the vacuum thickness to 12 Å. The energy cutoff is set to 500 eV. The Brillouin zone is sampled by a $12 \times 12 \times 1$ and $14 \times 14 \times 1$ k-mesh for 2D CrN and $CrB_2$, respectively. The second order force constants are computed with the density functional perturbation theory (DFPT) [47-49].

To search for the lowest energy structure of 2D $CrB_2$, we adopt our developed global optimization method. In our simulations, the total number of atoms is fixed to be no more than 12 in the unit cell (i.e. $CrB_2$, $Cr_2B_4$, $Cr_4B_8$). The magnetic state is set to be ferromagnetic (FM). In addition, we repeat the simulations by changing their magnetic state [i.e. antiferromagnetic (AFM) state] in order to make results reliable.

To estimate the Curie temperature of 2D CrN. We employ the parallel tempering Monte Carlo (PTMC) simulations [50,51]. A 80×80×1 supercell of the hexagonal unit cell is adopted. The number of replicas is set to 256.

***2D hyperferroelectric metal-***Since Shi *et al.* experimentally discovered the first "ferroelectric metal" in bulk $LiOsO_3$ in 2013 [52], people began to accept the view point that polar distortion can exist in a metal despite of the electron screening effects. Since the electric polarization of a 3D "ferroelectric metal" can not be switched by an electric field, we will refer to "ferroelectric metal" as polar metal hereafter. The schematic diagram of a three dimensional (3D) polar metal is shown in Fig. 1(a). It has a non-zero polarization in the bulk and possesses double well potential energy surface, but it cannot be switched by an electric field since the electrons are conductive. For 2D polar metals, the electric polarization could be either in-plane or out-of-plane, as schematically shown in Figs. 1(b) and 1(c), respectively. The 2D polar metal with an in-plane polarization (Fig. 1(b)) cannot be switched by an electric field similar to the case of 3D polar metal. Fig. 1(c) shows the case of a 2D hyperferroelectric metal with an out-of-plane polarization. Different from the other two kinds of polar metals (Figs. 1(a) and 1(b)), this kind of 2D hyperferroelectric metals (In general, the out-of-plane ferroelectricity can be either proper or improper) can be switched by a perpendicular electric field since electrons are confined within the slab and non-conductive along the out-of-plane direction. This is similar to the idea of switching the out-of-plane polarization of the thin-film of 3D polar metals with an electric field [53,54].

The new concept of 2D hyperferroelectric metals facilitates the search and design of 2D FM-FE multiferroics. This is because 2D FM metals are much easier to find than 2D FM insulators. Once the mechanism for stabilizing the out-of-plane polarization of

2D hyperferroelectric metals is identified, 2D FM-FE multiferroics can be achieved by introducing magnetism to the system.

***CrN as a 2D hyperferroelectric metal and a 2D multiferroic-***The previously proposed 2D AB binary monolayers [29] (A and B atoms belong to group IV or III-V) are 2D insulating hyperferroelectrics. Here, we proposed that 2D CrN hexagonal monolayer (buckled) is a 2D hyperferroelectric metal. Previous studies [55,56] have already showed that the 2D buckled CrN hexagonal monolayer adopts a FM metallic ground state. However, they did not investigate the microscopic mechanism for the FM ground state and the interesting FE-FM coupling in this system, which we will discuss later in detail. Here, based on first-principles calculations, we show that the buckled FE configuration of CrN (Fig. 2(a)) has a lower energy by 14 meV/f.u. than the paraelectric (PE) planar structure (Fig. 2(b)). Since the FE CrN is polar, we also use the dipole moment correction method to check the result. We find the FE CrN also has a lower energy by 12meV/f.u. than the planar CrN with the dipole moment correction method. Hence, the dipole moment correction affects little the result. The electric polarization of buckled FE CrN is about $6.2 \times 10^{-12}$ C/m. We use the nudge elastic band (NEB) method to find the transition path and estimate the energy barrier [see Fig. S1(a) of supporting information (SM)]. The phonon calculations indicate that FE CrN is dynamically stable (see Fig. S1(b) of SM). Our calculated band structures of PE and FE CrN show that 2D CrN is metallic (see Figs. 2(d) and (e)). To check whether the polarization in FE CrN can be switched by an electric field, we apply an electric field along the +z (0.7 V/Å) direction on the FE CrN with the initial polarization along −z to find that its polarization will be inverted (Fig. 2(c)) after structural relaxation. Experimentally, it is challenging to apply such a large electric field (0.7 V/Å). However, this simple simulation shows that 2D CrN is indeed a 2D hyperferroelectric metal. Note that it was suggested [55] that the 2D CrN can be grown on the $MoSe_2$ or $MoS_2$ substrates due to the small lattice mismatch".

Both PE and FE CrN adopt a FM spin ground state. Therefore, 2D CrN is a rare 2D FM-FE multiferroic. We note that although hydroxylized graphone was predicted to be a 2D FM-FE multiferroic, the magnetic moments are contributed by the dangling 2p electrons of carbon [16]. For the PE CrN, the total energy of the FM state is lower than the nonmagnetic (paramagnetic) state, collinear $2\times1$ AFM state and 120°-Néel AFM state (see Fig. S2(a)-(f) of SM), by 544 meV/Cr, 133 meV/Cr, 151 meV/Cr, respectively. For the FE CrN, the corresponding energy differences are 772 meV/Cr, 250 meV/Cr, 309 meV/Cr, respectively. From the energy difference between the FM and collinear $2\times1$ AFM state, the nearest neighboring Cr-Cr exchange parameters for PE and FE CrN are -33 meV and -63 meV, respectively (In this work, we use the effective exchange parameter by setting the spin value to 1). The Curie temperatures in the PE and FE CrN are estimated to be 420 K and 805 K, respectively (see Fig. S3(a), (b) of SM). Hence, we can see that the FM exchange interaction in FE CrN is stronger than that in PE CrN.

The stronger FM exchange interactions in FE CrN indicate that FE ordering enhances the FM ordering. On the other hand, we find that the FM ordering can also enhance the FE ordering. If we consider the paramagnetic nonmagnetic states, we find that the PE planar CrN has a lower energy than the FE buckled CrN. With the FM state, the optimized FE CrN structure has a lower energy than the optimized PE CrN structure by 14 meV/f.u.. However, with the collinear $2\times1$ AFM state, the FE CrN structure has a higher energy than the PE CrN structure by 104 meV/f.u.. These test calculations prove that FM ordering can enhance the FE ordering. We propose that this novel FM-FE coupling can be described by the spin-phonon coupling term $E = \lambda u_{1z} u_{2z} \vec{S}_1 \cdot \vec{S}_2$, where $\lambda$ is the coupling constant, $u_{iz}$ and $\vec{S}_i$ represent the FE displacements and spins of two neighboring Cr atoms, respectively. The form of this spin-phonon coupling is compatible with the $D_{3h}$ symmetry of the paramagnetic PE CrN. The coupling constant can be estimated by comparing the force constants of PE CrN between the FM state and the collinear $2\times1$ AFM state. The force constant between two neighboring Cr ions in the FM state along the z direction is -0.226 eV/Å². While for the two neighboring Cr ions with opposite spins in the collinear $2\times1$ AFM state, it is 0.093 eV/Å². Therefore, the coupling constant is estimated to be $\lambda = -0.16$ eV/Å². The negative sign of the coupling constant $\lambda$ suggests that the total energy will be lowered if both of the two neighboring Cr ions with parallel spins are displaced along the same out-of-plane direction. Hence, the out-of-plane ferroelectricity in 2D CrN is induced by an unusual spin-phonon coupling, which was shown to be also responsible for the strain induced FM-FE multiferroicity in SrMnO$_3$ [57]. We note that the hyperferroelectricity in 2D FM CrN is not due to the lone pair of N ions, different from the 2D P$_2$O$_3$ case [30]. In fact, the planar hexagonal structure has a lower energy than the FE hexagonal structure for AlN, ScN, and nonmagnetic CrN.

We now try to understand the microscopic physical mechanism of FM-FE coupling in 2D CrN. A clue can be seen from a comparison of the band structures of PE CrN and FE CrN. For the PE CrN in the FM state, the spin-majority $\{d_{xz}, d_{yz}\}$ band crosses with the $\{d_{x^2-y^2}, d_{xy}\}$ band since $\{d_{xz}, d_{yz}\}$ orbital and $\{d_{x^2-y^2}, d_{xy}\}$ orbital are odd and even with respect to the in-plane mirror plane, respectively. This is similar to the nodal line semimetal state in 2D nonmagnetic Hg$_3$As$_2$ system [58] except that here 2D CrN is magnetic. For the FE CrN, the spin-majority $\{d_{xz}, d_{yz}\}$ band now hybridizes with the $\{d_{x^2-y^2}, d_{xy}\}$ band to open a direct band gap, while the global indirect band gap still remains negative. The opening of a direct band gap leads to an energy lowering of FE CrN. It appears that the interaction between the $\{d_{xz}, d_{yz}\}$ orbital and the $\{d_{x^2-y^2}, d_{xy}\}$ orbital is the key to understand the FM-FE coupling. For 2D CrN, detailed local density of states (DOS) analysis (see Fig. S4 of SM) shows that the Cr$^{3+}$ ions take the high-spin electronic configuration (3 $\mu_B$), as shown in Fig. 3. The energies of the d-orbitals increase in the order of $d_{z^2}, \{d_{xz}, d_{yz}\}, \{d_{x^2-y^2}, d_{xy}\}$. Because of the band formation, the spin-majority $\{d_{xz}, d_{yz}\}$ and $\{d_{x^2-y^2}, d_{xy}\}$ levels are partially filled, resulting in the FM ground state. As can be seen in the lower-panel of Fig. 3(a), the spin-up $\{d_{xz}, d_{yz}\}$ ($\{d_{x^2-y^2}, d_{xy}\}$) orbitals of a Cr ion (Cr$_1$) can interact with the spin-up $\{d_{xz}, d_{yz}\}$ ($\{d_{x^2-y^2}, d_{xy}\}$) orbitals of a neighboring Cr ion (Cr$_2$) through the shared N atom (see Fig. 2(c)). This interaction will lead to low-lying bonding states and high-lying anti-bonding states. The occupation of the bonding states will result in an energy lowering of the FM state. The above FM mechanism can also apply to the case of FE CrN. But now the additional orbital interactions make the FM interaction stronger in FE CrN. This can be explained as following. For the PE CrN, the effective hopping between the $\{d_{xz},d_{yz}\}$ orbital of Cr$_1$ and the $\{d_{x^2-y^2},d_{xy}\}$ orbital of Cr$_2$ is zero by symmetry. In contrast, there is a non-zero effective hopping between the $\{d_{xz},d_{yz}\}$ orbital of Cr$_1$ and the $\{d_{x^2-y^2},d_{xy}\}$ orbital of Cr$_2$ in FE CrN. This is illustrated in Fig. 3(b). A spin up electron can hop from the $Cr_1:d_{xz}$ orbital to the nearest neighboring $N:p_x$ orbital since the buckling breaks the mirror symmetry, then to the $Cr_2:d_{x^2-y^2}$ orbital due to the p-d $\sigma$ interaction. Effectively, there is an interaction between $Cr_1:d_{xz}$ orbital and $Cr_2:d_{x^2-y^2}$ orbital. These additional interactions will further lower the energy of the bonding state, and enhance the FM interaction in FE CrN.

To confirm the orbital interactions [59] in 2D CrN, we construct the maximally-localized Wannier function (MLWF) [60-62] of the spin-majority channel. We choose all p orbitals of N atom and all d orbitals of Cr atom as the spin-majority projection functions. For PE CrN, we find the hopping parameter $t(Cr_1:d_{xz} \to N_{px})$ is zero. While for FE CrN, this becomes 1.25 eV because of the buckling. In PE CrN, the spin-majority $\{d_{xz},d_{yz}\}$ band crosses with the $\{d_{x^2-y^2},d_{xy}\}$ band (see Fig. 4(a)) For the FE CrN, the spin-majority $\{d_{xz},d_{yz}\}$ band now hybridizes with the $\{d_{x^2-y^2},d_{xy}\}$ band to open a direct band gap. This direct band gap opening can be explained by the non-zero hopping interaction $t(Cr_1:d_{xz} \to N_{px})$ in FE CrN (see Fig. 4(b)).

We also perform another MLWF analysis to direct calculate the effective hopping between the Cr d orbitals. In this case, we choose all d orbitals for Cr atoms as the spin-majority projection functions. For FE CrN, we find that the magnitudes of the effective hopping terms $t(d_{xz}^{Cr_1},d_{xy}^{Cr_2})$, $t(d_{xz}^{Cr_1},d_{x^2-y^2}^{Cr_2})$, $t(d_{yz}^{Cr_1},d_{xy}^{Cr_2})$ and $t(d_{yz}^{Cr_1},d_{x^2-y^2}^{Cr_2})$ are equal to 0.36 eV, 0.15 eV, 0.21 eV and 0.42 eV, respectively. For PE CrN, these effective hopping terms are exactly zero due to the in-plane mirror symmetry. This analysis supports the scenario shown in Fig. 3.

The above discussion shows that 2D CrN is a 2D FM-FE multiferroic, which can be used to realize novel nanoscale multiple state memory. Note that the switching of the FE polarization might not change the magnetic state of 2D CrN since the coupling between the polarization $\vec{P}$ and magnetization $\vec{M}$ is of the usual $P^2M^2$ type. Below we will show that it is possible to realize electric-field control of magnetism with the 2D hyperferroelectric metal.

***Electric-field control of magnetism with a 2D hyperferroelectric-***Our idea to realize electric field control of magnetism is illustrated in Fig. 5(a). The initial state (ground state) has zero out-of-plane electric polarization (antiferroelectric (AFE) or FE with an in-plane electric polarization, metal or insulator) with magnetic state-I, while the final state (metastable state) is a 2D hyperferroelectric (metal or insulator) with magnetic state-II. When one switches the states between the initial state and the final state with an electric field along the z direction, the magnetic property changes since the initial and final state are associated with different magnetic states. In this way, one realizes the electric field control of magnetism, similar to the concept of 3D asymmetric multiferroics [63]. It is noted that if the ground state of a system is FE and the metastable state is AFE, one may not realize the electric field control of magnetism since the electric field will switch between the two equivalent FE state with opposite polarizations and the higher energy AFE state might never be stabilized.

Recently, 2D $FeB_2$ was found to adopt an interesting buckled hexagonal structure ($C_{6v}$ point group) [64]. This structure has a lower energy than the planar structure (left panel of Fig. 5(b)). Thus, 2D $FeB_2$ can be considered as a 2D hyperferroelectric metal. Unfortunately, 2D FE $FeB_2$ is nonmagnetic, thus is not a multiferroic. With our developed global optimization methods [65,66] for searching 2D materials, we find the ground state of $CrB_2$ is a planar structure ($CrB_2$-$D_{2h}$, see left-panel of Fig. 5(b)). The metastable structure of $CrB_2$ is a 2D hyperferroelectric ($CrB_2$-$C_{6v}$, see right-panel of

Fig. 5(b)). The DOS calculations indicate that they are both metallic (see Fig. S5 of SM). These two structures have different magnetic ground states. For planar $CrB_2$-$D_{2h}$, it adopts an AFM magnetic ground state with the total energy lower than that of FM state by about 96 meV/f.u.. We use four different magnetic configurations (see Fig. S6 of SM) to estimate the exchange parameters $J_{12}$, $J_{13}$, and $J_{23}$ with Heisenberg model. Their values are estimated to be 90 meV, -35 meV and -35 meV, respectively. The strongest exchange interaction is the AFM interaction ($J_{12}$) between the two nearest neighbor Cr ions, which probably originates from the direct Cr-Cr exchange. The other Cr-Cr exchange interactions are weakly FM. For $CrB_2$-$C_{6v}$, the magnetic ground state is FM with the total magnetic moment about 1.5 $\mu_B$/f.u., which has a lower energy than the collinear $2\times1$ AFM state by 13 meV/Cr. The energy difference between AFM planar $CrB_2$-$D_{2h}$ and FM-FE $CrB_2$-$C_{6v}$ is about 397 meV/f.u.. We use the NEB method to estimate the energy barrier (Note that since the planar $CrB_2$-$D_{2h}$ and FE $CrB_2$-$C_{6v}$ adopt a different magnetic state, we use the FM state to estimate the energy barrier). The path for FM planar $CrB_2$-$D_{2h}$ and FM-FE $CrB_2$-$C_{6v}$ transition is shown in Fig. S7(a) of SM. The energy barrier (about 363 meV/Cr) is large, suggesting that it may not be switched easily by an external electric field. Note that in this work, we mainly focus on the idea of electric field controlling of magnetism with the use of a hyperferroelectric metal. The electric polarization of FM-FE $CrB_2$-$C_{6v}$ is calculated to be about $9.0\times10^{-13}$ C/m. The FM nature of the magnetic coupling in metallic $CrB_2$-$C_{6v}$ is understandable since the coexistence of ferromagnetism and metallicity is common. In addition, the phonon calculations indicate that both $CrB_2$-$D_{2h}$ and $CrB_2$-$C_{6v}$ are dynamically stable (see Figs. 12(b), (c)). Our results suggest that 2D $CrB_2$ might be adopted to realize electric field control of magnetism: A large out-of-plane electric field can switch the AFM planar state to FE-FM state, while a relatively smaller out-of-plane electric field will switch the FE-FM state back to the original AFM planar state.

Finally, we briefly discuss the origin of hyperferroelectricity in $MB_2$ (M represents transition metal) systems. The FE $MB_2$-$C_{6v}$ (M=Fe, Cr) structure is more stable than the PE planar $MB_2$-$D_{6h}$ (see Fig. S8(a) of SM) structure. In the FE $MB_2$ structure, the nearest neighbor M-M distance (3.16 Å for $CrB_2$) is smaller than that (3.48 Å for $CrB_2$) in the PE planar MB$_2$-D$_{6h}$. We also find that the FE MB$_2$ structure is more stable than the MB$_2$ structure with M atoms located at both sides of the boron plane (see Fig. S8(b) of SM). Therefore, the metal-metal bonding in MB$_2$ appears to be a new mechanism for stabilizing the out-of-plane electric polarization in 2D systems. For CrB$_2$, the planar CrB$_2$-D$_{2h}$ has a lower energy than the PE CrB$_2$-D$_{6h}$ structure since the Jahn-Teller like distortion helps to enhance some of the B-B and Cr-Cr bondings.

**Conclusion-**In summary, we have proposed a new concept of 2D hyperferrelectric metals and showed that 2D hyperferrelectric metals can be used to design 2D FM-FE multiferroics. These concepts are demonstrated by taking 2D CrN and CrB$_2$ as representative examples. In particular, 2D CrN is found to be a 2D hyperferroelectric metal and FM-FE multiferroic, while CrB$_2$ is a 2D asymmetric multiferroic suitable for realizing electric field control of magnetism. We reveal that spin-phonon coupling and metal-metal interactions are two new mechanisms to compete with the depolarization field for stabilizing the out-of-plane electric polarization. Our work not only extend the switchable FE to metallic systems, but also pave a new avenue to design/search 2D magnetoelectric multiferroics.

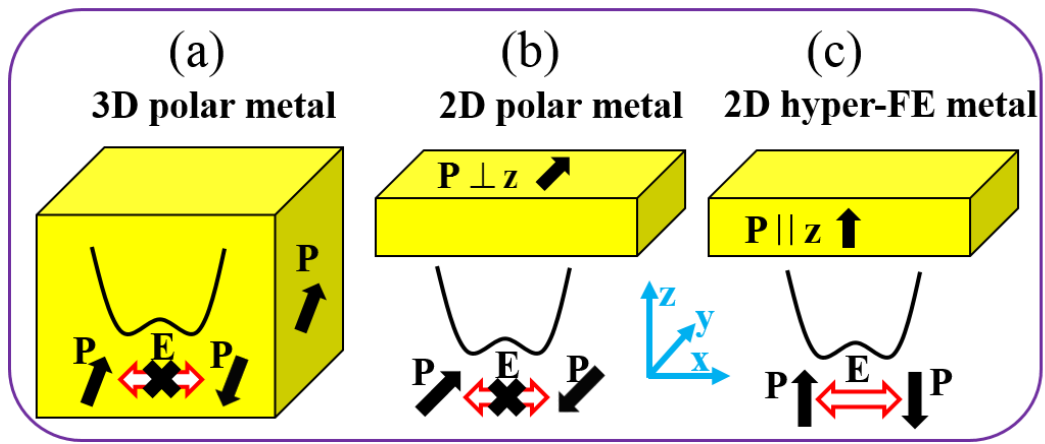

FIG. 1. (a) The schematic diagram of a 3D polar metal. It adopts a non-zero electric polarization in the bulk and possesses double well potential energy surface, but it cannot be switched by an electric field since electrons are conductive. (b) The schematic

diagram of a 2D polar metal with an in-plane polarization. Similar to the case of 3D polar metals, it cannot be switched by an electric field. (c) The schematic diagram of a 2D hyperferroelectric metal with an out-of-plane polarization. It can be switched since electrons are confined within the slab and non-conductive along the out-of-plane direction.

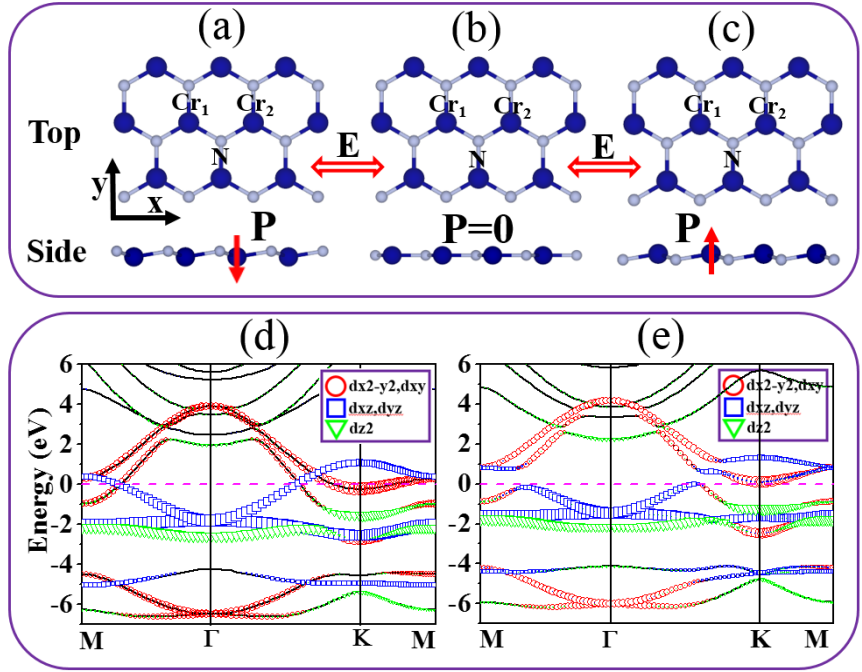

FIG. 2. (a), (b), (c) The crystal structure of FE (polarization along the –z, the buckling parameter is about 0.2 Å), PE, FE (polarization along the -z) CrN. The bluejkbjk and gray balls represent Cr and N atoms respectively. With an external perpendicular electric field, the polarization of FE CrN can be switched. (d), (e) The band structures of PE and FE CrN in the FM state. For PE CrN, the $\{d_{xz},d_{yz}\}$ and $\{d_{x^2-y^2},d_{xy}\}$ band form a nodal line since they adopt different eigenvalues for the in-plane mirror symmetry. However, for FE CrN, the $\{d_{xz},d_{yz}\}$ and $\{d_{x^2-y^2},d_{xy}\}$ band hybridize with each other as a result of a breaking of the in-plane mirror symmetry, leading to a band gap opening along the nodal line.

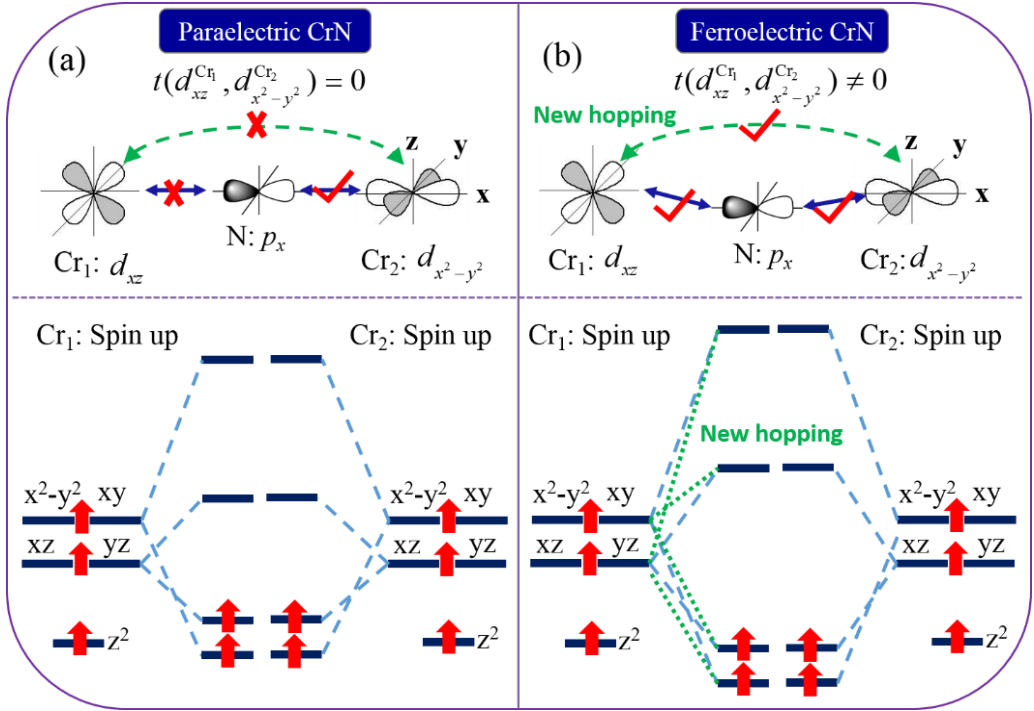

FIG. 3. Illustration of the microscopic physical mechanism of FM-FE coupling in the 2D CrN. (a) For PE CrN, the spin-up $\{d_{xz}, d_{yz}\}$ ($\{d_{x^2-y^2}, d_{xy}\}$) orbitals of a Cr ion (Cr$_1$) can interact with the spin-up $\{d_{xz}, d_{yz}\}$ ($\{d_{x^2-y^2}, d_{xy}\}$) orbitals of a neighboring Cr ion (Cr$_2$) through the shared N atom, lead to low-lying bonding states and high-lying anti-bonding states. The occupation of the bonding states lowers the total energies of the FM state. Due to the in-plane mirror symmetry, the $d_{xz}^{Cr_1}$ and $d_{x^2-y^2}^{Cr_2}$ orbitals can not hybridize with each other (i.e., $t(d_{xz}^{Cr_1}, d_{x^2-y^2}^{Cr_2}) = 0$). (b) For FE CrN, its FM mechanism is similar to that of PE CrN. However, the additional orbital interactions strength the FM interaction. For example, $t(d_{xz}^{Cr_1}, d_{x^2-y^2}^{Cr_2})$ becomes non-zero after the buckling breaks the in-plane mirror symmetry, so that the $d_{xz}^{Cr_1}$ and $d_{x^2-y^2}^{Cr_2}$ orbitals can hybridize with each other.

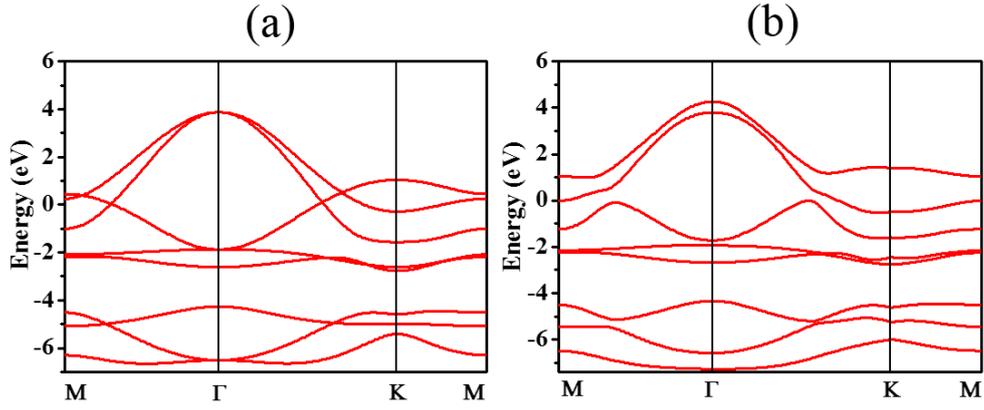

FIG. 4. (a) Spin-majority band structure of PE CrN calculated from the MLWF Hamiltonian. It is consistent with the DFT result [Fig. 2(e)]. (b) Spin-majority band structure computed with the MLWF Hamiltonian of FE CrN and an additional hopping [i.e., $t(Cr_1:d_{xz} \to N_{px})$]. It shows that the non-zero $t(Cr_1:d_{xz} \to N_{px})$ can reproduce the main feature of the band structure of FE CrN where the $\{d_{xz}, d_{yz}\}$ band anti-crosses with the $\{d_{x^2-y^2}, d_{xy}\}$ band.

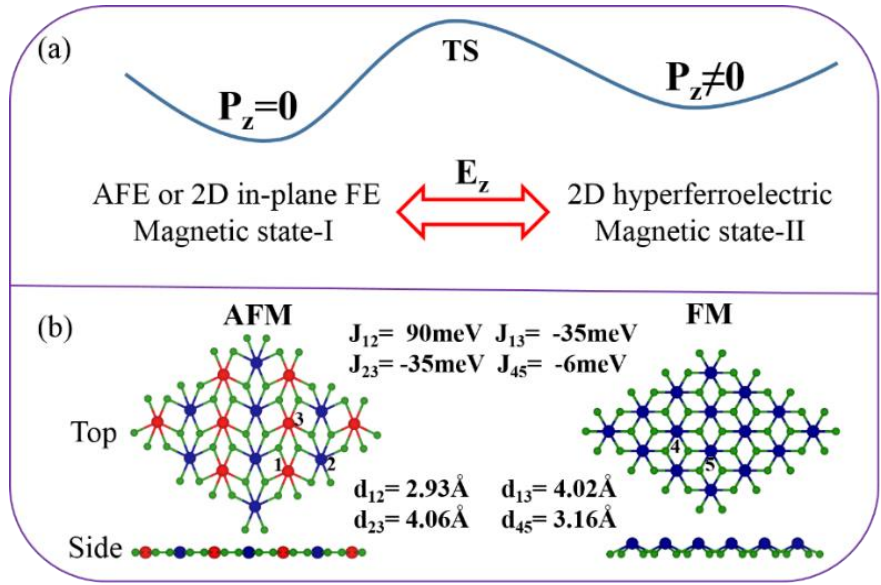

FIG. 5. (a) The schematic diagram of electric field control of magnetism with a 2D hyperferroelectric. The initial structure could be AFE or an in-plane FE, and the final structure is a 2D hyperferroelectric. These two states should have different magnetic ground states (magnetic state-I and magnetic state-II). (b) The crystal structure of CrB$_2$-

$D_{2h}$ (left-panel) and CrB$_2$-C$_{3v}$ (right-panel, the buckling parameter is about 0.9 Å). For CrB$_2$-D$_{2h}$, it adopts an AFM ground state. This is probably due to the direct AFM exchange interactions ($J_{12}$) between the two nearest neighbor Cr ions. The red and blue balls represent the spin up and spin down Cr ions, respectively. In contrast, CrB$_2$-C$_{3v}$ adopts a FM ground state. The average neighboring Cr-Cr bond length (3.16 Å) in CrB$_2$-C$_{3v}$ is shorter than that (3.67 Å) in CrB$_2$-D$_{2h}$, suggesting that there is a stronger metal-metal interaction in CrB$_2$-C$_{3v}$.

**Acknowledgments**

We thank K. Liu for valuable discussions. Work was supported by NSFC, FANEDD, NCET-10-0351, Research Program of Shanghai Municipality and MOE, the Special Funds for Major State Basic Research, Program for Professor of Special Appointment (Eastern Scholar), and Fok Ying Tung Education Foundation.

**References**

[1] K. S. Novoselov, A. K. Geim, S. V. Morozov, D. Jiang, Y. Zhang, S. V. Dubonos, I. V. Grigorieva, and A. A. Firsov, Electric field effect in atomically thin carbon films, Science **306**, 666 (2004).

[2] L. Li, Y. Yu, G. J. Ye, Q. Ge, X. Ou, H. Wu, D. Feng, X. H. Chen, and Y. Zhang, Black phosphorus field-effect transistors, Nat. Nanotechnol. **9**, 372 (2014).

[3] Q. H. Wang, K. Kalantar-Zadeh, A. Kis, J. N. Coleman, and M. S. Strano, Electronics and optoelectronics of two-dimensional transition metal dichalcogenides, Nat. Nanotechnol. **7**, 699 (2012).

[4] M. Chhowalla, H. S. Shin, G. Eda, L.-J. Li, K. P. Loh, and H. Zhang, The chemistry of two-dimensional layered transition metal dichalcogenide nanosheets, Nat. Chem. **5**, 263 (2013).

[5] T. Hu, H. Wu, H. Zeng, K. Deng, and E. Kan, New Ferroelectric Phase in Atomic-Thick Phosphorene Nanoribbons: Existence of in-Plane Electric Polarization, Nano Lett. **16**, 8015 (2016).

[6] R. Fei, W. Kang, and L. Yang, Ferroelectricity and Phase Transitions in Monolayer

Group-IV Monochalcogenides, Phys. Rev. Lett. **117**, 097601 (2016).

[7] A. Lebedev, Quasi-two-dimensional ferroelectricity in KNbO3/KTaO3 superlattices, Phys. Solid State **53**, 2463 (2011).

[8] W. Ding, J. Zhu, Z. Wang, Y. Gao, D. Xiao, Y. Gu, Z. Zhang, and W. Zhu, Prediction of intrinsic two-dimensional ferroelectrics in In2Se3 and other III2-VI3 van der Waals materials, Nat. Commun. **8**, 14956 (2017).

[9] M. Mehboudi, B. M. Fregoso, Y. Yang, W. Zhu, A. van der Zande, J. Ferrer, L. Bellaiche, P. Kumar, and S. Barraza-Lopez, Structural phase transition and material properties of few-layer monochalcogenides, Phys. Rev. Lett. **117**, 246802 (2016).

[10] M. Mehboudi, A. M. Dorio, W. Zhu, A. van der Zande, H. O. Churchill, A. A. Pacheco-Sanjuan, E. O. Harriss, P. Kumar, and S. Barraza-Lopez, Two-Dimensional Disorder in Black Phosphorus and Monochalcogenide Monolayers, Nano Lett. **16**, 1704 (2016).

[11] M. T. Ong and E. J. Reed, Engineered piezoelectricity in graphene, ACS nano **6**, 1387 (2012).

[12] M. Noor-A-Alam and Y.-H. Shin, Switchable polarization in an unzipped graphene oxide monolayer, Phys. Chem. Chem. Phys. **18**, 20443 (2016).

[13] W. Li and J. Li, Ferroelasticity and domain physics in two-dimensional transition metal dichalcogenide monolayers, Nat. Commun. **7**, 10843 (2016).

[14] L. Kou, Y. Ma, C. Tang, Z. Sun, A. Du, and C. Chen, Auxetic and Ferroelastic Borophane: A Novel 2D Material with Negative Possion's Ratio and Switchable Dirac Transport Channels, Nano Lett. **16**, 7910 (2016).

[15] M. Wu and X. C. Zeng, Intrinsic Ferroelasticity and/or Multiferroicity in Two-Dimensional Phosphorene and Phosphorene Analogues, Nano Lett. **16**, 3236 (2016).

[16] M. Wu, J. D. Burton, E. Y. Tsymbal, X. C. Zeng, and P. Jena, Hydroxyl-decorated graphene systems as candidates for organic metal-free ferroelectrics, multiferroics, and high-performance proton battery cathode materials, Phys. Rev. B **87**, 081406 (2013).

[17] H. Wang, X. Li, J. Sun, Z. Liu, and J. Yang, BP_5 Monolayer with Multiferroicity and Negative Poisson's Ratio: A Prediction by Global Optimization Method, 2D Mater. **4**, 045020 (2017)

[18] H. Wang and X. Qian, Two-dimensional multiferroics in monolayer group IV monochalcogenides, 2D Mater. **4**, 015042 (2017).

[19] V. Garcia, S. Fusil, K. Bouzehouane, S. Enouz-Vedrenne, N. D. Mathur, A. Barthelemy, and M. Bibes, Giant tunnel electroresistance for non-destructive readout of ferroelectric states, Nature **460**, 81 (2009).

[20] R. Guo, L. You, Y. Zhou, Z. S. Lim, X. Zou, L. Chen, R. Ramesh, and J. Wang, Non-volatile memory based on the ferroelectric photovoltaic effect, Nat. Commun. **4**, **1990** (2013).

[21] J. F. Scott, [3D] Nano-Scale Ferroelectric Devices for Memory Applications, Ferroelectrics **314**, 207 (2005).

[22] K. Chang *et al.*, Discovery of robust in-plane ferroelectricity in atomic-thick SnTe, Science **353**, 274 (2016).

[23] M. Wu, S. Dong, K. Yao, J. Liu, and X. C. Zeng, Ferroelectricity in Covalently functionalized Two-dimensional Materials: Integration of High-mobility Semiconductors and Nonvolatile Memory, Nano Lett. **16**, 7309 (2016).

[24] I. Batra, P. Wurfel, and B. Silverman, New type of first-order phase transition in ferroelectric thin films, Phys. Rev. Lett. **30**, 384 (1973).

[25] W. Zhong, R. King-Smith, and D. Vanderbilt, Giant LO-TO splittings in perovskite ferroelectrics, Phys. Rev. Lett. **72**, 3618 (1994).

[26] M. Dawber, K. Rabe, and J. Scott, Physics of thin-film ferroelectric oxides, Rev. Mod. Phys. **77**, 1083 (2005).

[27] S. N. Shirodkar and U. V. Waghmare, Emergence of ferroelectricity at a metal-semiconductor transition in a 1T monolayer of MoS2, Phys. Rev. Lett. **112**, 157601 (2014).

[28] A. Chandrasekaran, A. Mishra, and A. K. Singh, Ferroelectricity, Antiferroelectricity, and Ultrathin 2D Electron/Hole Gas in Multifunctional Monolayer MXene, Nano Lett. **17**, 3290 (2017).

[29] D. Di Sante, A. Stroppa, P. Barone, M.-H. Whangbo, and S. Picozzi, Emergence of ferroelectricity and spin-valley properties in two-dimensional honeycomb binary compounds, Phys. Rev. B **91**, 161401 (2015).

[30] W. Luo and H. Xiang, Two-Dimensional Phosphorus Oxides as Energy and Information Materials, Angew. Chem. Int. Ed. **128**, 8717 (2016).

[31] K. F. Garrity, K. M. Rabe, and D. Vanderbilt, Hyperferroelectrics: proper ferroelectrics with persistent polarization, Phys. Rev. Lett. **112**, 127601 (2014).

[32] P. Li, X. Ren, G. C. Guo, and L. He, The origin of hyperferroelectricity in LiBO3 (B = V, Nb, Ta, Os), Sci Rep **6**, 34085 (2016).

[33] L. Seixas, A. Rodin, A. Carvalho, and A. C. Neto, Multiferroic Two-Dimensional Materials, Phys. Rev. Lett. **116**, 206803 (2016).

[34] S.-W. Cheong and M. Mostovoy, Multiferroics: a magnetic twist for ferroelectricity, Nat. Mater. **6**, 13 (2007).

[35] S. Picozzi and C. Ederer, First principles studies of multiferroic materials, J. Phys. Condens. Matter **21**, 303201 (2009).

[36] D. Khomskii, Trend: Classifying multiferroics: Mechanisms and effects, Physics **2**, 20 (2009).

[37] Y. Tokura and S. Seki, Multiferroics with spiral spin orders, Adv. Mater. **22**, 1554 (2010).

[38] J. M. Rondinelli and C. J. Fennie, Ferroelectricity: Octahedral Rotation‐Induced Ferroelectricity in Cation Ordered Perovskites (Adv. Mater. 15/2012), Adv. Mater. **24**, 1918 (2012).

[39] H. J. Zhao, W. Ren, Y. Yang, J. Íñiguez, X. M. Chen, and L. Bellaiche, Near room-temperature multiferroic materials with tunable ferromagnetic and electrical properties, Nat. Commun. **5**, 4021 (2014).

[40] S. Callori, J. Gabel, D. Su, J. Sinsheimer, M. Fernandez-Serra, and M. Dawber, Ferroelectric PbTiO 3/SrRuO 3 superlattices with broken inversion symmetry, Phys. Rev. Lett. **109**, 067601 (2012).

[41] Z. Tu, M. Wu, and X. C. Zeng, Two-Dimensional Metal-Free Organic Multiferroic Material for Design of Multifunctional Integrated Circuits, J. Phys. Chem. Lett. **8**, 1973 (2017).

[42] N. A. Hill, Why are there so few magnetic ferroelectrics?, J. Phys. Chem. B **104**, 6694 (2000).

[43] G. Kresse and J. Furthmüller, Efficient iterative schemes for ab initio total-energy calculations using a plane-wave basis set, Phys. Rev. B **54**, 11169 (1996).

[44] P. E. Blöchl, Projector augmented-wave method, Phys. Rev. B **50**, 17953 (1994).

[45] G. Kresse and D. Joubert, From ultrasoft pseudopotentials to the projector augmented-wave method, Phys. Rev. B **59**, 1758 (1999).

[46] J. P. Perdew, K. Burke, and M. Ernzerhof, Generalized gradient approximation made simple, Phys. Rev. Lett. **77**, 3865 (1996).

[47] S. Baroni, S. De Gironcoli, A. Dal Corso, and P. Giannozzi, Phonons and related crystal properties from density-functional perturbation theory, Rev. Mod. Phys. **73**, 515 (2001).

[48] X. Gonze, First-principles responses of solids to atomic displacements and homogeneous electric fields: Implementation of a conjugate-gradient algorithm, Phys. Rev. B **55**, 10337 (1997).

[49] X. Gonze and C. Lee, Dynamical matrices, Born effective charges, dielectric permittivity tensors, and interatomic force constants from density-functional perturbation theory, Phys. Rev. B **55**, 10355 (1997).

[50] K. Hukushima and K. Nemoto, Exchange Monte Carlo method and application to spin glass simulations, J. Phys. Soc. Jpn. **65**, 1604 (1996).

[51] P. Wang and H. Xiang, Room-temperature ferrimagnet with frustrated antiferroelectricity: promising candidate toward multiple-state memory, Phys. Rev. X **4**, 011035 (2014).

[52] Y. Shi *et al.*, A ferroelectric-like structural transition in a metal, Nat. Mater. **12**, 1024 (2013).

[53] H. Xiang, Origin of polar distortion in LiNbO 3-type "ferroelectric" metals: role of A-site instability and short-range interactions, Phys. Rev. B **90**, 094108 (2014).

[54] A. Filippetti, V. Fiorentini, F. Ricci, P. Delugas, and J. Iniguez, Prediction of a native ferroelectric metal, Nat Commun **7**, 11211 (2016).

[55] A. V. Kuklin, A. A. Kuzubov, E. A. Kovaleva, N. S. Mikhaleva, F. N. Tomilin, H. Lee, and P. V. Avramov, Two-dimensional hexagonal CrN with promising magnetic and optical properties: A theoretical prediction, Nanoscale **9**, 621 (2017).

[56] S. Zhang, Y. Li, T. Zhao, and Q. Wang, Robust ferromagnetism in monolayer chromium nitride, Sci. Rep. **4**, 5241 (2014).

[57] J. H. Lee and K. M. Rabe, Epitaxial-strain-induced multiferroicity in SrMnO3 from first principles, Phys. Rev. Lett. **104**, 207204 (2010).

[58] J. Lu, W. Luo, X. Li, S. Yang, J. Cao, X. Gong, and H. Xiang, Two-Dimensional Node-Line Semimetals in a Honeycomb-Kagome Lattice, Chin. Phys. Lett. **34**, 057302 (2017).

[59] John B. Goodenough, "Magnetism and the chemical bond" (1963).

[60] A. A. Mostofi, J. R. Yates, Y.-S. Lee, I. Souza, D. Vanderbilt, and N. Marzari, wannier90: A tool for obtaining maximally-localised Wannier functions, Comput. Phys. Commun. **178**, 685 (2008).

[61] N. Marzari and D. Vanderbilt, Maximally localized generalized Wannier functions for composite energy bands, Phys. Rev. B **56**, 12847 (1997).

[62] I. Souza, N. Marzari, and D. Vanderbilt, Maximally localized Wannier functions for entangled energy bands, Phys. Rev. B **65**, 035109 (2001).

[63] X. Lu and H. Xiang, Designing asymmetric multiferroics with strong magnetoelectric coupling, Phys. Rev. B **90**, 104409 (2014).

[64] H. Zhang, Y. Li, J. Hou, A. Du, and Z. Chen, Dirac state in the FeB2 monolayer with graphene-like boron sheet, Nano Lett. **16**, 6124 (2016).

[65] Y. Wang, J. Lv, L. Zhu, and Y. Ma, Crystal structure prediction via particle-swarm optimization, Phys. Rev. B **82**, 094116 (2010).

[66] W. Luo, Y. Ma, X. Gong, and H. Xiang, Prediction of silicon-based layered structures for optoelectronic applications, J. Am. Chem. Soc. **136**, 15992 (2014).